\documentstyle[12pt]{article}
\input epsf
\title{On Mean-Field Theory of Quantum Phase Transition in Granular 
Superconductors}
\author{M.V. Simkin \\ {\em Department of Physics, Brown 
University,}\\
{\em Providence, RI 02912-1843}}
\date{ }
\begin{document}
\maketitle
\begin{abstract}
In previous work on quantum phase transition in granular superconductors,
where mean-field theory was used, an assumption was made that the order
parameter as a function of the mean field is a convex up function.
Though this is not always the case in phase transitions, this assumption
must be verified, what is done in this article.
\end{abstract}

Quantum phase transition in granular superconductors, modeled as
Josephson - junction arrays (JJA) have been a subject of extensive
investigations. For a comprehensive review see a book by \v{S}im\'{a}nek
\cite{sh}. The picture of the phenomenon is as follows. At the bulk
transition temperature $T_0$ the magnitude of the order parameter
of each superconducting grain becomes nonzero. When the energy of
Josephson coupling between grains $E_J$ is less than $T_0$, thermal
fluctuations cause the phases $\theta$ of the order parameter on
different grains to be uncorrelated until the temperature is lowered
to $T=E_J$. When the charging energy $E_C$, i.e. the electrostatic 
energy of a Cooper-pair located on a fixed grain is comparable
to  $E_J$, the zero-point fluctuations of $\theta$ destroy the 
long-range
superconductive order even at zero temperature.

A lot of work have been done to study the phase diagram of the granalular
superconductor in ($T/E_C,E_J/E_C$) plane \cite{sh}. Mostly mean field
theory has been used, though recently a renormalization-group study
was employed \cite{gr}. In previous work, using mean-field theory, an
important assumption was made: that the order parameter 
$<\cos\theta>$ is a convex up function of effective mean field.
The aim of this article is to verify this assumption.

Quantum JJA are described by a Hamiltonian \cite{sh}:
\begin{equation}
	\hat{H}=\sum_{ik}P_{ik}(2e)^2\hat{n}_i\hat{n}_k-
\sum_{ik}E_{ik}\cos(\theta_i-\theta_k).
\end{equation}
Here matrix elements $P_{ik}$ are Coulomb interactions and $E_{ik}$ are
the Josephson energies between the $i$ and $k$ grains, 
$\hat{n}_k=-id/d\theta_k$
is the excess Cooper-pair number operator, conjugate to the phase 
$\theta_k$ of the $k$ grain. While $P_{ik}$ are nonzero for all pairs,
$E_{ik}$ are nonzero only for nearest neighbors. In the following we will
consider a periodic array of identical grains with identical Josephson
junctions between them. This idealization is reasonable when the disorder
in Josephson energies and Coulomb interactions is not to big. In opposite
case one faces a percolation problem.

The mean-field Hamiltonian in Hartree approximation is obtained by
replacing all operators, except two conjugate operators corresponding
to chosen grain, by their average values: 
\begin{equation}
	\hat{H}_{MF}=P(2e)^2\hat{n}^2-z\mu E_J\cos\theta.
\end{equation}
Here $z$ is the lattice coordination number; $P=P_{ii}$ is the diagonal
element of the potential matrix; $\mu$ is the quantum-statistical and spatial
average value
of $\cos\theta$; the average value of $\hat{n}$ is zero because of 
electroneutrality of a sample, and this is why the mean-field Hamiltonian
depends on diagonal elements of potential matrix only.

By introducing variables
\begin{equation}
	E_C=P(2e)^2 ;q=z\mu E_J/E_C,
\end{equation}
Eq.2 may be rewritten as
\begin{equation}
	\hat{H}_{MF}=-E_C(d^2/d\theta^2+q\cos\theta).
\end{equation}
We denote the eigenfunctions of the Hamiltonian in Eq.(4) as
$\Psi_n^1$ and $\Psi_n^2$ and its eigenvalues as $E_n^1$ and $E_n^2$.
We classify the eigenfunctions by condition $\Psi_n^1(q=0)=\cos(n\theta)$,
$\Psi_n^2(q=0)=\sin(n\theta)$.

For theory to be self-consistent the value of the order parameter,
obtained using the mean-field Hamiltonian should be equal to $\mu$:
\begin{equation}
	\mu=\frac{\sum_{n,\alpha} \exp(-E_n^\alpha/T)<\Psi_n^\alpha
	\mid \cos\theta \mid \Psi_n^\alpha>}{\sum_{n,\alpha} 
	\exp(-E_n^\alpha/T)}.
\end{equation}

From the Feynman-Hellman theorem one obtains
\begin{equation}
	<\Psi_n \mid \cos\theta \mid \Psi_n>=-\frac{1}{E_C}dE_n/dq.
\end{equation}
This means that we just need to calculate $E_n^1$ and $E_n^2$ to solve
the problem.
Hamiltonian in Eq.(4) is similar to the operator of the Mathieu equation 
\cite{mcl}:
\begin{equation}
	\hat{H}_{Mathieu}=-d^2/dx^2+2q\cos(2x).
\end{equation}
The relation between the eigen-values of (4) and (7) is:
\begin{equation}
	E_n^1(q)=E_C\frac{1}{4}a_{2n}(2q);
	E_n^2(q)=E_C\frac{1}{4}b_{2n}(2q).
\end{equation}

In Fig. 1 the order parameter $\mu$, as a function of the mean-field $q$,
calculated using Eqs. (5),(6),(8) is plotted. The eigenvalues of Mathieu
equation $a_{m}(q)$, $b_{m}(q)$ are calculated using a routine from the 
IMSL library. $m$'s up to ten have been taken into account.
For Fig.1(a) only even order Mathieu functions have been used. This 
corresponds to taking into account only $2\pi$-periodic $\Psi_n$, i.e. those
with integer $n$.
For Fig.1(b) both even and odd order Mathieu functions have been used,
corresponding to $4\pi$-periodic eigenfunctions of (4). Inclusion of 
$4\pi$-periodic eigenfunctions   may serve as
 an approximation to the continuous spectrum, i.e. to taking into account
all  $\Psi_n$ with real $n$. 
The latter situation corresponds to the case when the $2\pi$ periodicity
is broken by coupling of the Josephson-junction to an environment (for
example by normal shunt resistance),  which permits a continuous change
of the charge on the junction.

	\begin{figure}[htb]
	\centering
	\begin{minipage}{12.0cm}
	\epsfxsize=12.0cm
	\epsfbox[57   191  546   590] {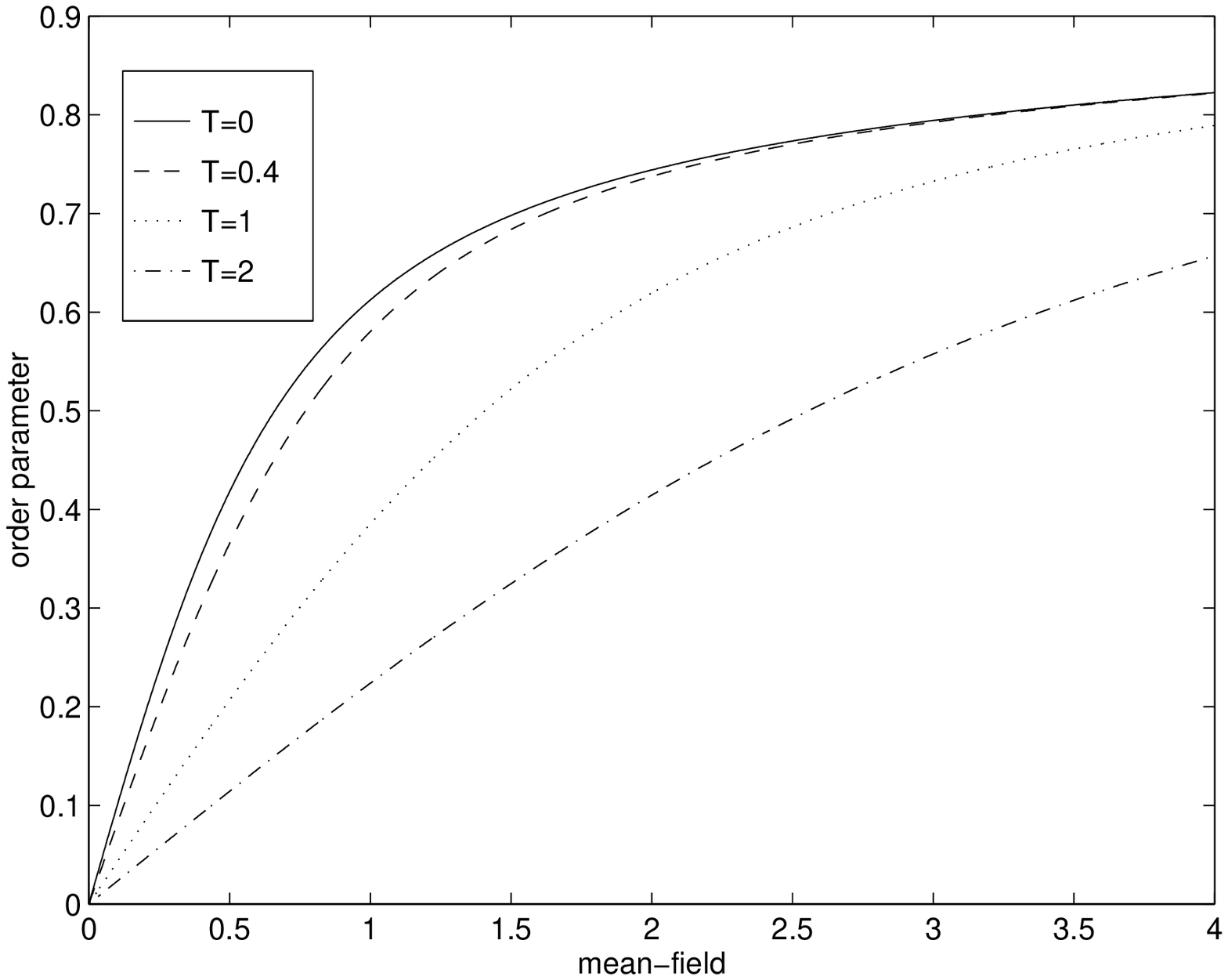}
	\end{minipage}
	\begin{minipage}{12.0cm}
	\epsfxsize=12.0cm
	\epsfbox[57   191   546   590]{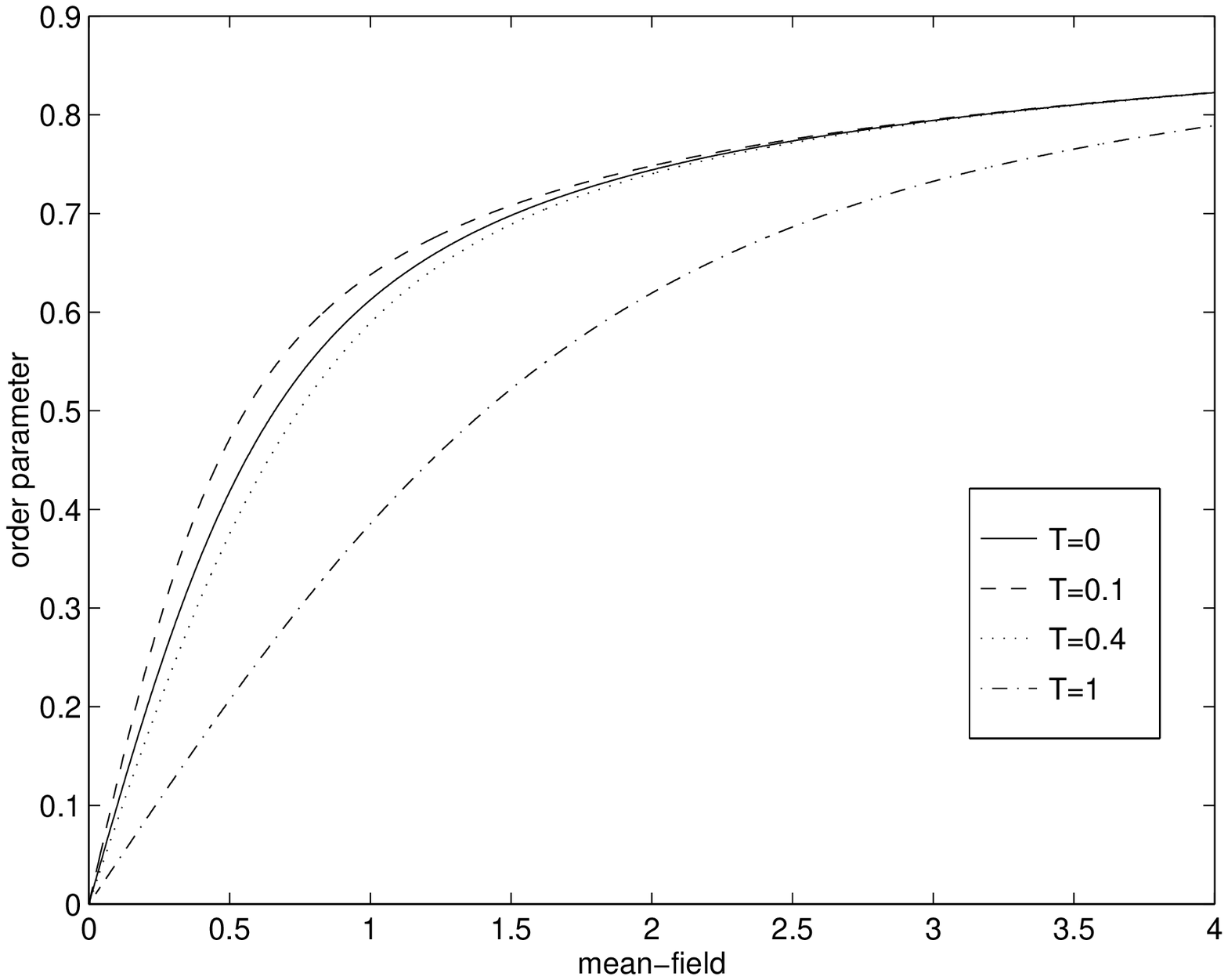}
	\end{minipage}
	\caption{Order parameter $\mu$ as a function of the mean-field 
	$q=z\mu E_J/E_C$. In (a) only $2\pi$ periodic eigenfunctions of the
	mean-field Hamiltonian were used. In (b) also $4\pi$ periodic ones
	were included.}
	\end{figure}

In the previous work only slope of $\mu(q)$ for small $q$ was calculated
using either expansion of Mathieu functions for $q \ll 1$ \cite{sh},
or perturbation series for (4) \cite{sim}.
Then superconducting transition temperature was calculated from:
\begin{equation}
	d\mu(q,T)/dq|_{q=0}=E_C/zE_J,
\end{equation}
assuming that $\mu(q)$ is convex up. Fig.1 proves that this assumption
is correct for both cases when only $2\pi$ and when also $4\pi$ periodic
eigenfunctions are included.

One can consider two types of transition. First is at fixed ratio $E_J/E_C$
and variable temperature. One can express the order
parameter $\mu$ through the mean field $q$ using Eq.3: $\mu=q/(zE_J/E_C)$,
and plot this line in Fig.1. The intersection of this line with different 
curves $\mu(q,T)$
gives the value of the order parameter for each temperature.
The second transition is at fixed temperature and variable ratio  $E_J/E_C$.
Here one should plot several lines $\mu=q/(zE_J/E_C)$ for different $E_J/E_C$,
and find their intersection with the fixed curve $\mu(q,T)$.
As all curves $\mu(q,T)$ in Fig.1 are convex up both mentioned above 
transitions are continuous.

In Fig.2 a phase diagram of the granular superconductor, calculated
using Eq. (9) is presented. In addition the curve for continuous spectrum
from Ref.\cite{sim} is given. A common feature of phase diagrams, obtained
with $4\pi$-periodic eigen-functions and with continuous spectrum is
a reentrant transition. Indications for such a transition were observed 
experimentally, for example, in Ref.\cite{vdz}.
	
	\begin{figure}[htb]
	\centering	
	\begin{minipage}{15.0cm}
	\epsfxsize=15.0cm
	\epsfbox[57   197   593   590] {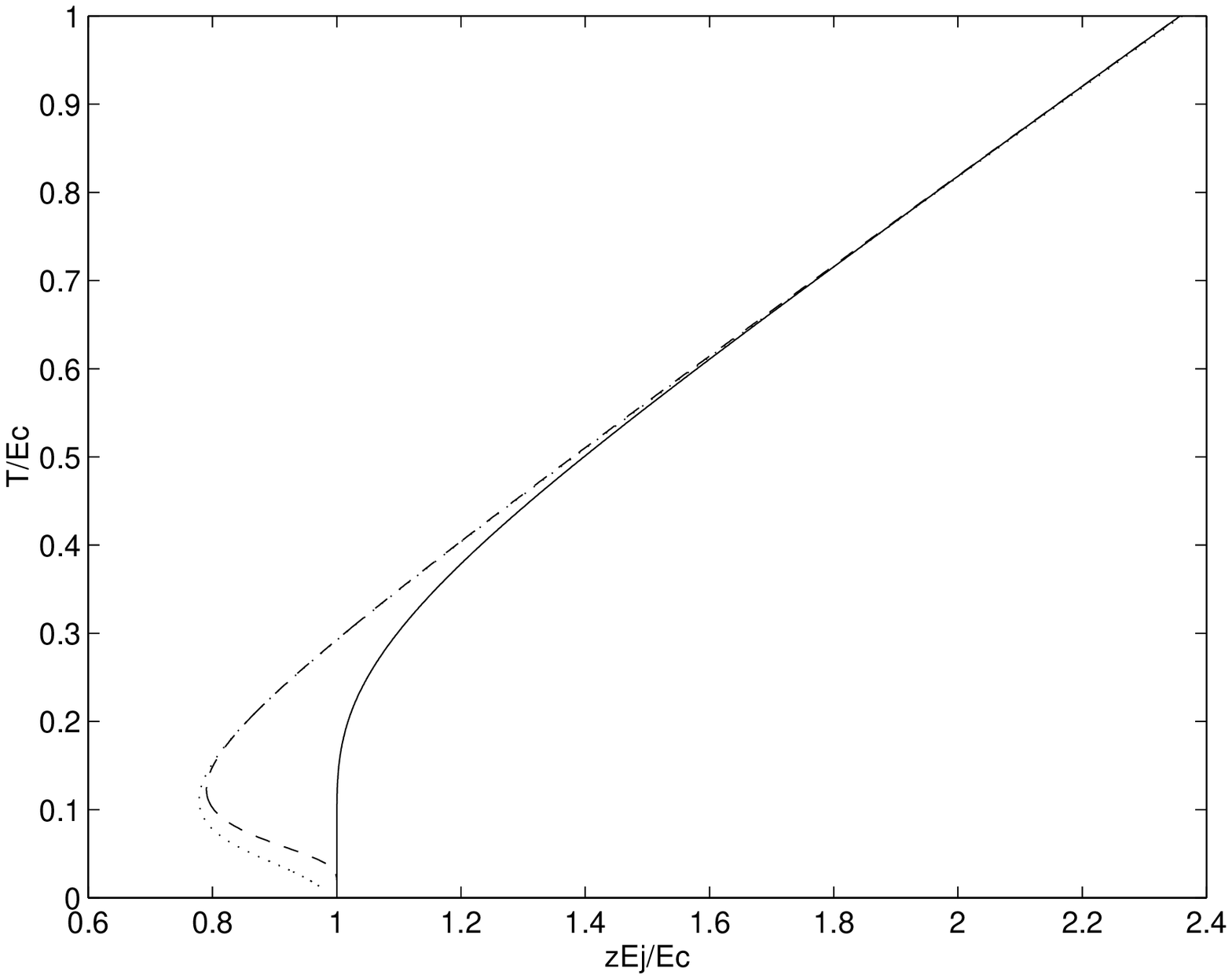}
	\end{minipage}
	\caption{The mean-field phase diagram of a granular superconductor.
	Solid line is obtained taking into account only $2\pi$-periodic 
	eigenfunctions of (4). Dashed line obtained by including also 
	$4\pi$-periodic
	ones. Dotted line is from Ref.[4], where a continuum of states was
	used.}
	\end{figure}

In conclusion, I have verified the assumption, made in earlier work,
that the order parameter of granular superconductor is a convex up
function of the mean-field, or in other words that the transition is of the
second order.

I am grateful to  V.K Ignatovich and J.M. Kosterlitz, for useful 
conversations and to E. \v{S}im\'{a}nek for correspondence. This work was 
supported by National Science Foundation Grant No. DMR-9222812.

\end{document}